\documentclass[12pt,preprint]{aastex}
\usepackage{graphicx,rotate,url,mathptmx,lscape,times}

\def\gtsim{\mathrel{\hbox{\rlap{\hbox{\lower4pt\hbox{$\sim$}}}\hbox{$>$}}}}
\def\lesssim{\mathrel{\hbox{\rlap{\hbox{\lower4pt\hbox{$\sim$}}}\hbox{$<$}}}}

\def\cm{{\rm\thinspace cm}}

\def\erg{{\rm\thinspace erg}}

\def\km{{\rm\thinspace km}}

\def\ps{{\rm\thinspace s^{-1}}}
\def\pcc{{\rm\thinspace cm^{-3}}}

\def\s{{\rm\thinspace s}}

\def\ergps{\mbox{$\erg\s^{-1}\,$}}

\def\kmps{\mbox{$\km\ps\,$}}

\def\pscm{\mbox{$\cm^{-2}\,$}}

\def\pscm{\mbox{$\cm^{-2}\,$}}
\def\pscmps{\mbox{$\cm^{-2}\, \s^{-1}\,$}}

\def\la{\mbox{{\rm L}$\alpha$}}

\def\hb{\mbox{{\rm H}$\beta$}}

\def\civ{\mbox{{\rm C~{\sc iv}}}}

\def\feii{\mbox{{\rm Fe~{\sc ii}}}}
\def\hplus{\mbox{{\rm H}$^+$}}

\def\h0{\mbox{{\rm H}$^0$}}
\def\he0{\mbox{{\rm He}$^0$}}

\DeclareMathAlphabet{\vib}{OML}{cmm}{m}{it}

\usepackage{aas_macros}

\shorttitle{Infalling \feii\ clouds in AGN}
\shortauthors{Ferland et al.}

\begin{document}

\title{Implications of infalling \feii - emitting clouds in active galactic
nuclei: anisotropic 
properties\footnote{Contains material \copyright\ British Crown copyright
2009/MoD.}}

\author{Gary J. Ferland\altaffilmark{2},
        Chen Hu\altaffilmark{3},
	Jian-Min Wang\altaffilmark{3,4},
	Jack A. Baldwin\altaffilmark{5},
	Ryan L. Porter\altaffilmark{6},
	Peter A. M. van Hoof\altaffilmark{7}
	and R.J.R. Williams\altaffilmark{8}}

\altaffiltext{2}{Department of Physics, University of Kentucky, Lexington,
KY 40506, USA.}

\altaffiltext{3}{Key Laboratory for Particle Astrophysics,
Institute of High Energy Physics, Chinese Academy of Sciences,
Beijing 100049, China.}

\altaffiltext{4}{Theoretical Physics Center for Science Facilities (TPCSF),
Chinese Academy of Sciences, China.}

\altaffiltext{5}{Department of Physics and Astronomy,
Michigan State University, Lansing, MI, USA.}

\altaffiltext{6}{Astronomy, University of Michigan, Ann Arbor, MI 48109, USA}

\altaffiltext{7}{Royal Observatory of Belgium, Ringlaan 3, 1180 Brussels,
Belgium}

\altaffiltext{8}{AWE plc, Aldermaston, Reading RG7 4PR, UK}


\begin{abstract}

\noindent
We investigate consequences of the discovery that
\feii\ emission in quasars, one of the spectroscopic signatures of
``Eigenvector 1'', may originate in infalling clouds.
Eigenvector 1 correlates with the Eddington ratio
$L / L_{Edd}$ so that \feii / \hb\ increases
as $L / L_{Edd}$ increases.
We show that the ``force multiplier'', the ratio of gas opacity
to electron scattering opacity, is $\sim 10^3 - 10^4$ in
\feii\ emitting gas.
Such gas would be accelerated away from the central
object if the radiation force is able to act on the entire cloud.
As had previously been deduced, infall requires that the clouds have large column densities
so that a substantial amount of shielded gas is present.
The critical column density required for infall to occur depends on
$L / L_{Edd}$, establishing a link between
Eigenvector 1 and the \feii /\hb\ ratio.
We see predominantly the shielded
face of the infalling clouds rather than the symmetric distribution
of emitters that has been assumed.
The \feii\ spectrum emitted by the shielded face
is in good agreement with observations thus solving
several long-standing mysteries in quasar emission lines.

\end{abstract}


\section{Introduction}

Principal component analysis (PCA) of the emission-line
spectra of quasars isolates several linearly independent components
(\citealp{1992ApJS...80..109B}).
These are ascribed to distinct physical processes within these objects.
Eigenvector 1, which is responsible for the largest part of the variability
within the samples, anti-correlates with $L / L_{Edd}$,
the ratio of the luminosity to the Eddington limit,
while Eigenvector 2 represents the accretion rate
\citep{2002ApJ...565...78B}.
It has been proposed that Eigenvector 1 represents the effects of radiation
pressure controlling the column densities of the surviving line-emitting
clouds \citep{marconi08,netzer09,dong09}.
This Letter is a specific, quantitative exploration of that idea.

\feii\ emission is one of the main spectroscopic signatures of
Eigenvector 1.
The physics of \feii\ emission is complex
with many levels contributing to the spectrum, which
occurs in a blended complex of muddled features
(\citealp{1985ApJ...288...94W},
\citealp{1999ApJS..120..101V},
\citealp{2004ApJ...615..610B}).
There has been only partial success in reproducing
the observed \feii\ emission in a realistic model of AGN emission-line regions
\citep{2004ApJ...615..610B}.

\citet{2009arXiv0908.0386G} has recently discussed the considerable range 
of existing observational evidence indicating that QSO emission-line regions 
include infalling gas. 
\citet{2008ApJ...683L.115H} and \citet{2008ApJ...687...78H} recently 
found that \feii\ emission is systematically redshifted relative 
to the QSO systemic velocities, and presented the reasons suggesting 
that this is because the \feii\ emission comes from infalling clouds. 
In a forthcoming paper (C. Hu et al. 2009, in  
preparation), we will further develop a model outlined by 
\citet{2008ApJ...687...78H} in which the \feii\ emission comes from 
infalling clouds systematically viewed from  
their shielded faces. Such clouds must have a large enough
column density to fall toward the black hole, despite
$L / L_{Edd} \sim 10^{-1} - 1$.
Here, we describe a preliminary exploration of the  
basic spectroscopic properties of clouds viewed from their non-illuminated
sides, which is an important question in its own right  independent of the
exact dynamical model. If indeed the \feii\ emission 
does come from infalling clouds, our results have major consequences 
for the predicted
emission-line spectrum and reveal one of the underlying
drivers for Eigenvector 1.

\section{Observed properties of \feii\ emitters}\label{s:obsfeii}

\citet{2008ApJ...683L.115H} and \citet{2008ApJ...687...78H}
find that much of the optical \feii\ emission comes from a
redshifted intermediate-line region.
The redshift of the \feii -emitting region inversely correlates with
the $L/L_{Edd}$ ratio as indicated by Eigenvector 1.
The correlations are that, as the $L/L_{Edd}$ ratio increases,
\feii / \hb\ increases  (\citealp{2002ApJ...565...78B},
\citealp{1992ApJS...80..109B})
and the \feii\ redshift relative to
the systemic rest frame decreases (\citealp{2008ApJ...687...78H}).
These relations are the subject of this Letter.

For simplicity, we consider two \feii\ emission
bands. The optical \feii\ $\lambda 4558$ band is the integrated \feii\
emission between $\lambda 4434$ and $\lambda 4684$, as defined by
\citet{1992ApJ...398..476F}.
UV \feii\ emission forms a blended pseudo continuum
with very few isolated \feii\ lines.
We follow the \citet{2004ApJ...615..610B} definition of
UV \feii\ $\lambda 2445$
as the integrated \feii\ flux
over the wavelength range $\lambda \lambda 2240 - 2650$.

We consider the properties of a typical
\feii -emitting cloud.
The observed spectrum is actually produced
by a mix of clouds with a broad range of densities and
distances from the central object
\citep{1995ApJ...455L.119B}.
However, in this preliminary investigation, we consider only a single cloud.

We use a combination of line-continuum reverberation
studies and theoretical predictions to determine
the parameters of this typical cloud.
\citet{1993ApJ...404..576M}
found that the UV \feii\ lines in NGC 5548 respond  to the driving
continuum on
a timescale similar to that of \civ\ $\lambda$1549 and somewhat
shorter than that of H$\beta$, but the optical \feii\ lines
in this object are too weak for useful reverberation measurements
\citep{2005ApJ...625..688V}.
\citet{2008ApJ...673...69K} studied the reverberation behavior of
the optical
\feii\ lines in Akn 120, for which there are no UV reverberation
measurements but which has strong optical \feii\ lines.
They found that in Akn 120 the optical \feii\ emission clearly does
not originate in the same region as \hb, and that there was
some evidence of
a reverberation response time of 300 days which implies
an origin
in a region several times further away from the central object than H$\beta$.
A 300 day
light-travel time corresponds to a source-cloud separation of
$\sim 8 \times 10^{17} \cm$.
For a general AGN spectral energy distribution (SED), the quoted luminosity of
Akn 120 ($10^{45} \ergps$), and this separation,
the flux of hydrogen-ionizing photons,
$\Phi (\mathrm{H} )$,
at the derived position of the \feii\ emission
is in the neighborhood of
$\Phi (\mathrm{H}) \approx 10^{19} \pscmps$.
From the results shown in Figure 7 of
\citet{2004ApJ...615..610B}, we estimate the gas density to be
$n( \mathrm{H} ) \sim 10^{10} \pcc$ and the turbulence
to be $u_{turb} \sim 10^2 \kmps$.
The ``standard cloud'' considered in the remainder of this Letter
has
these quantities and solar abundances.

\section{Calculations}

All calculations were done with version C10 of the spectral simulation code Cloudy,
last described by \citet{1998PASP..110..761F}.
We concentrate on the emission
from the shielded face of the cloud.
Cloudy has long predicted the fraction of the emission
that is beamed toward the central object \citep{1992ApJ...387...95F},
but the outward emission, which is often quite faint for the
large column densities we consider below, has not been an emphasis.
The inward versus outward \feii\ emission, although calculated, was not
reported.
We have improved the adaptive logic used to define the spatial grid
to better resolve the emission from the shielded face and added
several output options to report these predictions.
We now explicitly predict the inward, outward,
and total \feii\ emission in the form presented
by \citet{2004ApJ...615..610B}.

We consider the ionizing flux versus gas density plane presented by
\citet{2004ApJ...615..610B}.
For an isotropic radiation field, the ionizing flux $\Phi (\mathrm{H}) \propto r^{-2}$
so this variable can be thought of as a proxy for the
continuum source-cloud separation.
All possible AGN clouds lie somewhere on this plane.
Regions with the same ionization parameter form parallel
lines with a slope of unity.
Little observable emission comes from the large flux, small density quadrant
which corresponds to highly ionized gas with temperatures near the
Compton limit \citep{1997ApJS..108..401K}.
Clouds in the small flux, large density quadrant have very low ionization,
tend to be cool, and produce strong \feii\ emission.

\subsection{The force multiplier}

The fact that the \feii-emitting clouds, with $L/L_{Edd} \sim
10^{-1}$, are falling into the center has major consequences for the
cloud column density.  The Eddington limit is the luminosity at which
the force due to electron scattering balances the gravitational
acceleration for optically thin gas,
\begin{equation}
\label{eqn:EddingtonLimit}
L_{Edd}  = \frac{4\pi GMm_p c}{\sigma _T }.
\end{equation}
Here $\sigma _T$ is the Thomson cross section.  The ratio $L/L_{Edd}$
is a dimensionless measurement of the specific luminosity of the AGN.

\begin{figure}
\centering
\includegraphics[width=0.75\textwidth]{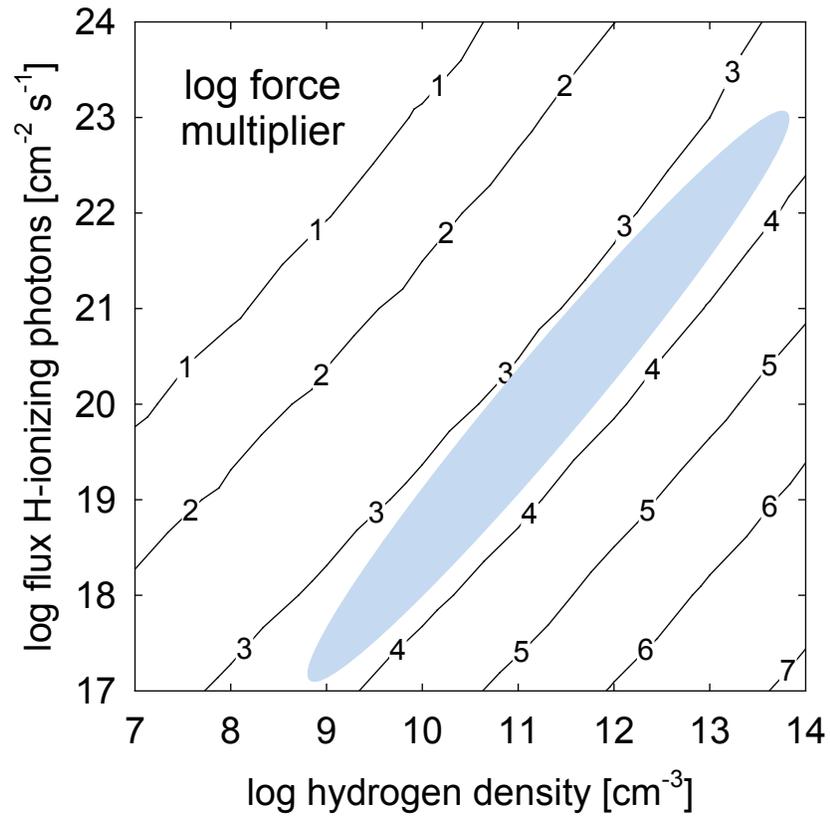}
\caption{Force multiplier, evaluated at the illuminated face of the cloud,
over the photon flux-density plane.  The shaded ellipse is the region capable of reproducing \feii\
observed emission \citep{2004ApJ...615..610B}.
\label{fig:ForceMultiplier}
}
\end{figure}

This acceleration limit is appropriate for very highly ionized gas.
For gas with a higher total cross section
$\sigma$, the outward force is greater by a
force multiplier $\sigma/\sigma_T$.  In general, $\sigma \gg \sigma _T$
for low-ionization gas due to the added opacity caused by line and
photoelectric absorption (\citealp{1974MNRAS.169..279C,1974A&A....34..211K,
1975ApJ...195..157C}).  The force multiplier over the range of
densities and ionizing fluxes is shown in Figure
\ref{fig:ForceMultiplier}, which also shows, as a shaded ellipse, the region
that \citet{2004ApJ...615..610B} found could produce acceptable \feii\
emission.

Figure \ref{fig:ForceMultiplier} shows that typical \feii-emitting
clouds should be strongly accelerated away from the central object
unless the Eddington ratio is below $L/L_{Edd} \sim 10^{-3} -
10^{-4}$.  The objects in the \citet{2008ApJ...687...78H} sample have
$L/L_{Edd} \sim 10^{-1}$ so the gas would be expected to be
accelerated outward.  How can it be in a state of infall?

\subsection{The need for large column densities}

The outward radiation pressure corresponding to the force multiplier
shown in Figure \ref{fig:ForceMultiplier} will act only on the
illuminated face of a large column density cloud.  This is because the
peak of the SED emitted by the AGN occurs at
ionizing energies.  The photons most capable of pushing the gas away
ionize the gas and are absorbed.

An \hplus--\h0\ ionization front occurs when most ionizing photons
have been absorbed.  Most of the outward momentum acts on the \hplus\
layer.  Relatively little energy is in high-energy photons which
penetrate into neutral regions.  The result is that most of the
radiative acceleration occurs in the \hplus\ layer which then pushes
on deeper neutral and molecular regions.
The presence of shielded regions allows inflow to occur even at super-Eddington
luminosities, as has been discussed previously
(e.g., \citealp{1998ApJ...494L.193S}). \citet{marconi08} and
\citet{netzer09} have previously discussed radiation pressure from ionizing
photons acting on large column-density AGN clouds, in terms of the effect on
the deduced black hole masses and the $L/L_{\rm Edd}$ versus $M_{\rm BH}$
relation.

\begin{figure}
\centering
\includegraphics[angle=-90,width=0.85\textwidth]{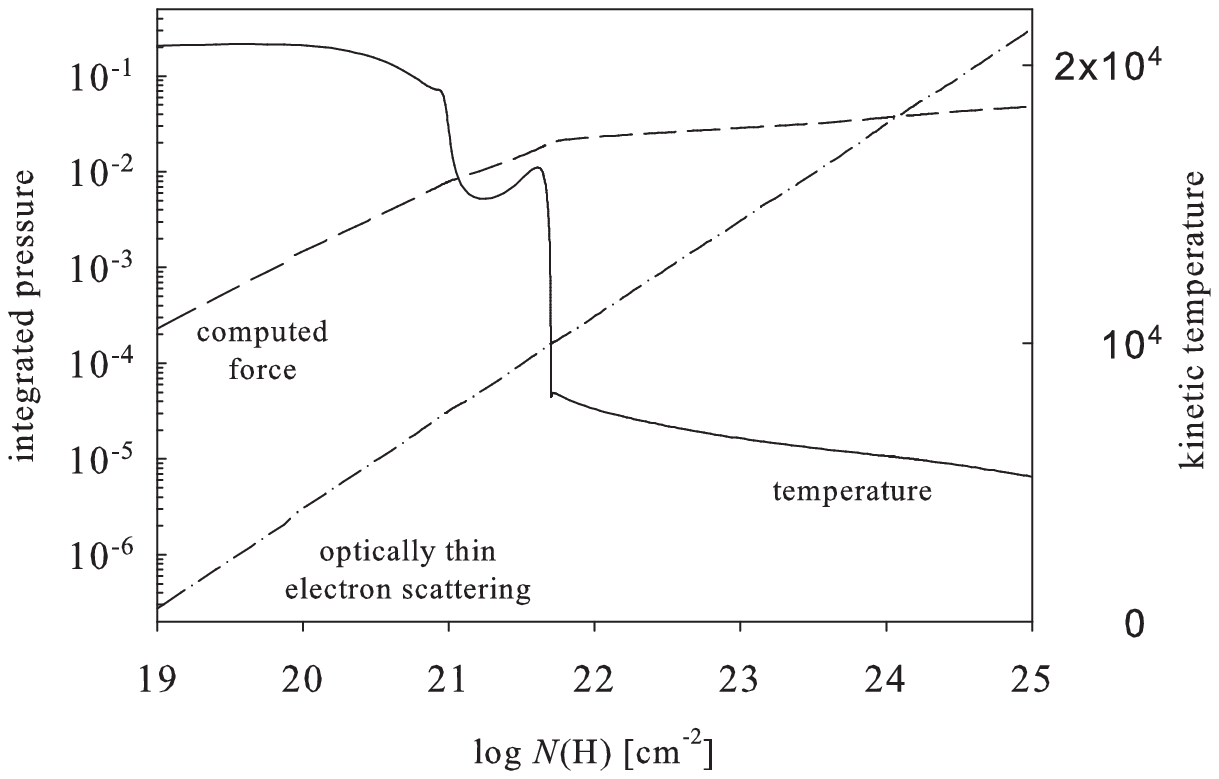}
\caption{Gas kinetic temperature, the computed force for the
radiative acceleration, and the
force for optically thin electron scattering, are shown
as a function of column density from the illuminated face.
Ionizing radiation enters from the left.
The two lines showing the integrated force intercept at a column density of
$N(\mathrm{H}) \approx 1.2\times 10^{24} \pscm$, the point where the
cloud would have neutral buoyancy for an AGN at the Eddington Limit.
\label{fig:RadiativeAccelerationDepth}
}
\end{figure}

We quantify this with a calculation shown in Figure
\ref{fig:RadiativeAccelerationDepth}, using the standard conditions described in Section~\ref{s:obsfeii}.
The gas kinetic temperature is shown as a function
of column density from the illuminated face of the cloud.
A sharp drop in temperature occurs at the \hplus--\h0\
ionization front and
an \hplus\ layer with a column density
of $N(\mathrm{H}) \approx 5 \times 10^{21} \pscm$ exists
on the face of the cloud.
The vast majority of the cloud consists of warm atomic gas
where the \feii\ lines form.

The dashed line shows the
computed radiative acceleration.
Within the \hplus\ layer
the force increases nearly linearly with increasing
column density as more of the incident radiation field is absorbed.
The force has nearly reached its asymptotic value when the
\hplus--\h0 ionization front is reached.
This ionized layer then pushes against
the much larger column density of cooler atomic gas.
The atomic layer adds mass to the cloud without contributing
to its outward acceleration.

In practice, the clouds may be subject to Rayleigh-Taylor instability
(\citealp{1982ApJ...252...39M}, \citealp{1986ApJ...305..187M}).  While
the radiation force will suppress instability at the front surface of
the cloud, it will enhance the instability at the ionization front.
Within the ionized region, the structure may be subject to additional
radiation-driven instabilities
(\citealp{1976A&A....52..203M,2000MNRAS.316..803W}), as well as
significant structural perturbations due to the varying continuum.
These and other sources may support a level of turbulence similar to
that which we model.  We will not address the stability of the clouds
we model further.  The model does require that the cloud integrity
must be maintained by other means, possibly magnetic, or, more likely,
that the cloud is an evolving dynamic entity that survives long enough to
be accelerated by the gravitational forces in the region.

Figure \ref{fig:RadiativeAccelerationDepth}
establishes the minimum column density where infall is possible.
The most useful way to consider the results in Figure \ref{fig:RadiativeAccelerationDepth}
is by reference to the acceleration that would occur for
pure electron scattering opacity in fully ionized gas since that is the opacity
used in the definition of the Eddington limit.
The dash-dotted line
shows the equivalent radiative driving for the case where
only electron scattering opacity acting on the incident
radiation field is considered.
The actual acceleration, the dashed line, is considerably larger than
the electron scattering acceleration across the \hplus\ region,
as expected from Figure \ref{fig:ForceMultiplier}.
The acceleration increases very slowly after the
\hplus -- \h0\ ionization front, while, with our definitions,
the reference electron scattering acceleration increases
linearly with column density $N(\mathrm{H})$.
The two lines cross at a column density of
$N( {\rm{H}} ) \approx 1.2 \times 10^{24} \pscm$,
the point where the cloud would be neutrally buoyant for
$L/L_{Edd} = 1$.

The minimum column density for which the inward gravitational force is
stronger than the total outward radiation pressure on a cloud,
$N(\mathrm{H})_{\rm infall}$, may be estimated as
\begin{equation}
N\left( {\rm{H}} \right)_{\rm infall}  \ge
{f_{\rm thick}\over \sigma_T } \frac{L}{L_{Edd}}
\simeq 1.5 \times 10^{24}f_{\rm thick}\frac{L}{L_{Edd}}{\rm\,cm^{ - 2}},
\label{eqn:ColDenEddRatio}
\end{equation}
\noindent
where $f_{\rm thick} \sim 1$ is the fraction of the incident radiation to
which the cloud is optically thick.  This establishes a relation
between specific luminosity and the minimal column density of
infalling clouds.  In objects of higher specific luminosity, which
tend to have stronger \feii\ emission, only high column density clouds
will be able to fall inward.

\begin{figure}
\centering
\includegraphics[angle=-90,width=0.85\textwidth]{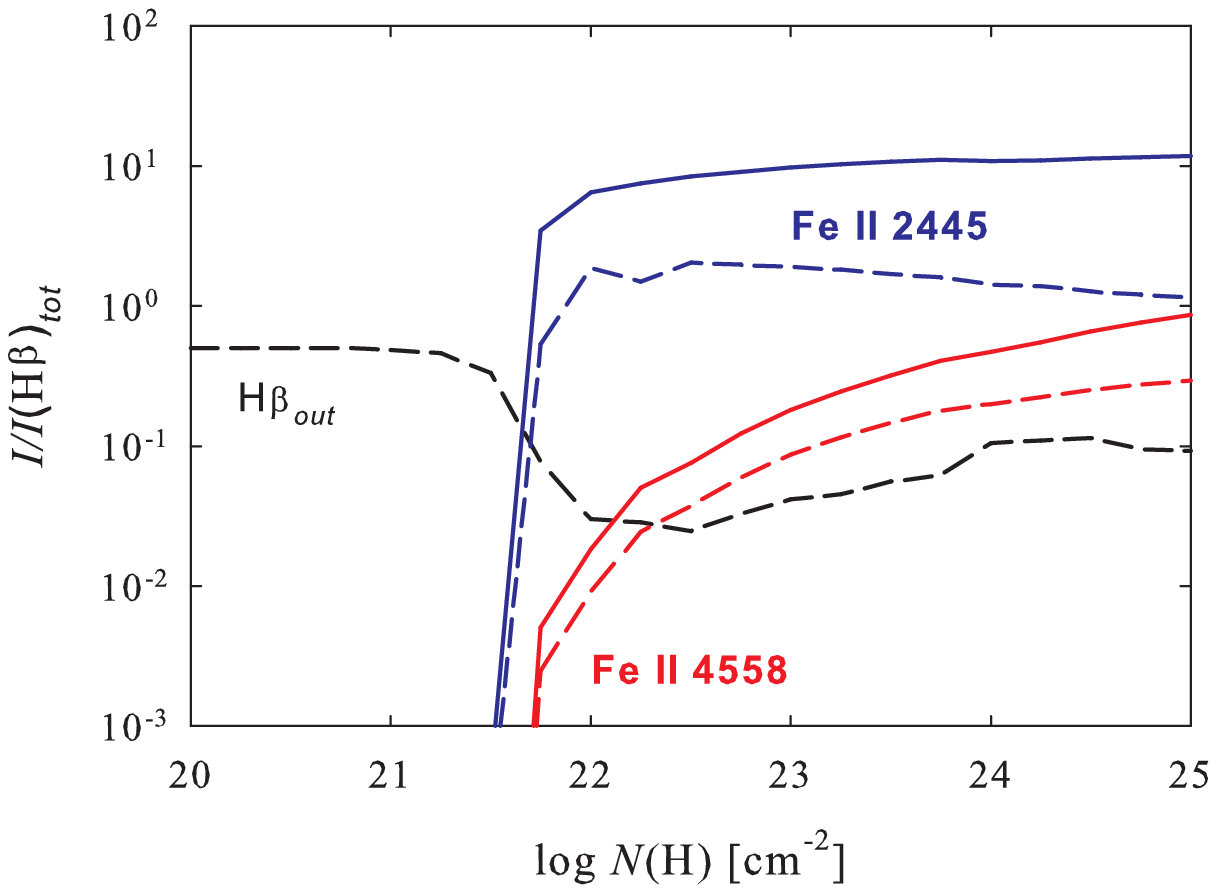}
\caption{Predicted total (solid line) and outward (dashed line)
emission relative to the total \hb\ is shown as a function of column
density.  \hb\ is isotropically emitted for small column densities
while the optical \feii\ band remains nearly isotropic for all column
densities.  The UV \feii\ band is predominantly inwardly beamed due to
the larger optical depth in these lines.  }
\label{fig:VaryColumnDensity}
\end{figure}

Figure \ref{fig:VaryColumnDensity} shows how the total and, separately, 
the outward emission varies as the cloud column density is increased. 
The \feii\ lines begin to form when clouds have column densities
$N(\mathrm{H}) > 10^{21.7} \pscm$, the point where the
$\hplus$--$\h0$ ionization front occurs.
This is also the column density where \hb\ becomes inwardly beamed.
\la\ is nearly fully emitted in the inward direction
\citep{1979ApJ...229..274F}.

\subsection{Isotropy of optical \feii\ emission}

\begin{figure}
\centering
\includegraphics[width=0.75\textwidth]{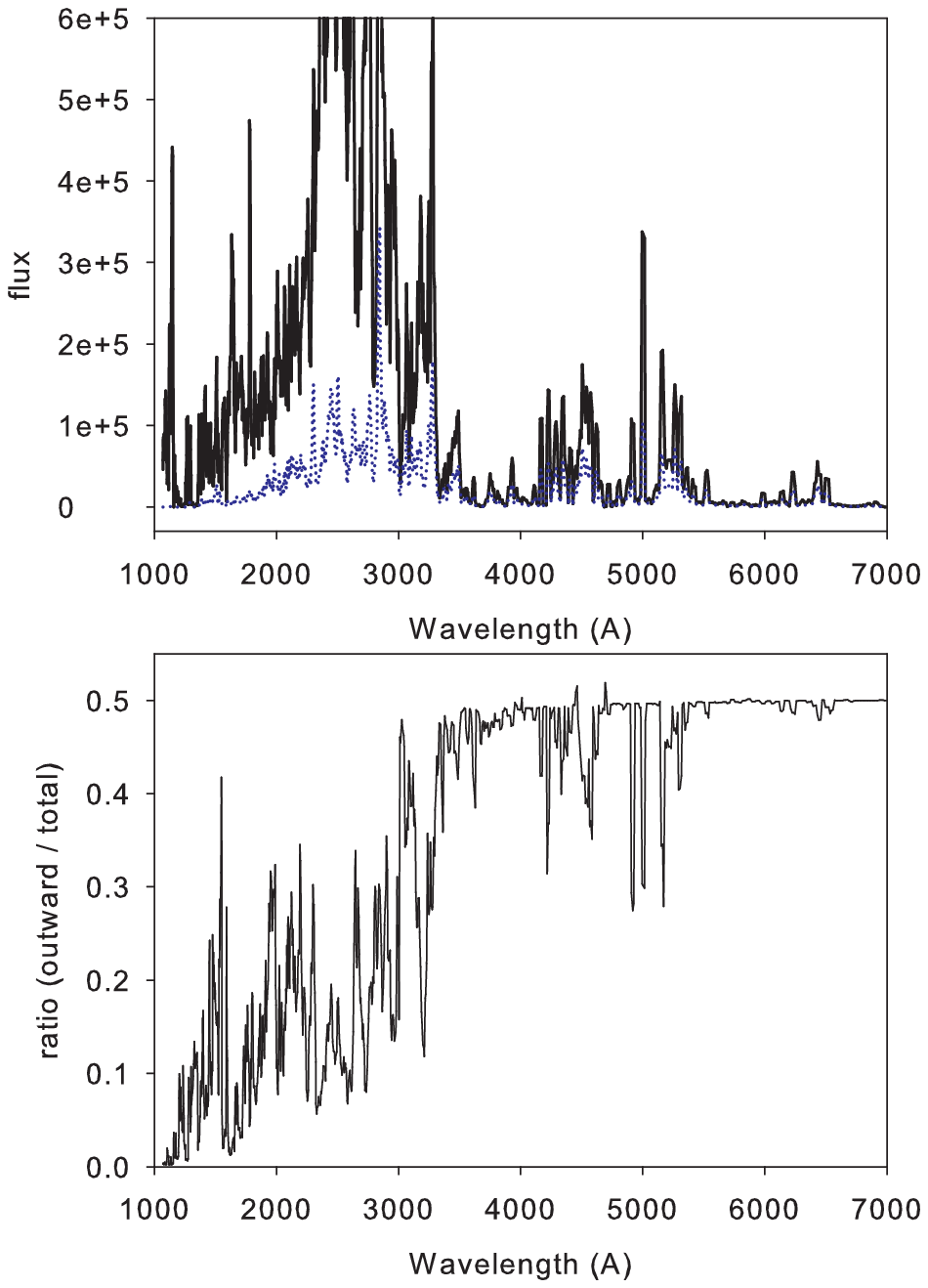}
\caption{Upper panel compares the total (solid) and outward (dotted)
\feii\ emission of our standard cloud.
The lower panel shows the ratio of the outward to total emission.
\label{fig:spectrum}
}
\end{figure}

The upper panel of Figure \ref{fig:spectrum} shows the total and outward \feii\ emission for our standard cloud.
The spectrum is complex and it is hard to distinguish between
the total and outward components in certain parts.
The lower panel shows the ratio of
the outward to the total emission.
This shows the general trend for the inward fraction to increase
with decreasing wavelength, mainly due to the larger optical
depths of UV \feii\ transitions.

\begin{figure}
\centering
\includegraphics[angle=-90,width=\textwidth]{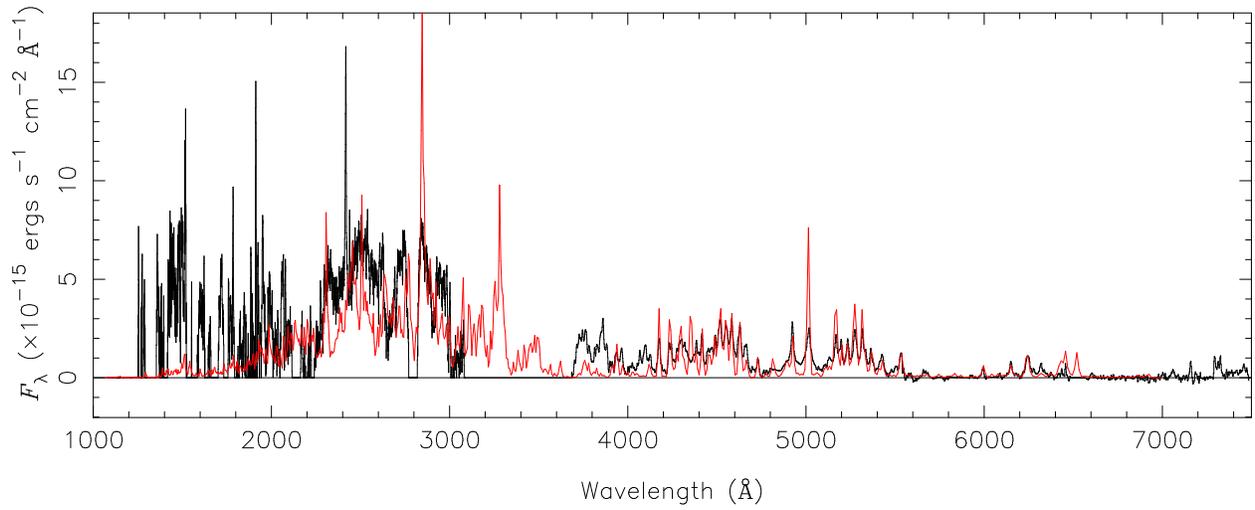}
\caption{Red smoother line is the outward \feii\ emission
from the standard model broadened to a FWHM of $900 \kmps$  while
the black spiky line is the observed \feii\ template.
Intermediate-line region clouds account for the optical emission
while allowing part of the UV emission to come from the inner broad-line region.
\label{fig:FeIICompareObserved}
}
\end{figure}

Figure \ref{fig:FeIICompareObserved} compares the outward emission
from our standard model with the \feii\ template presented by
\citet{2001ApJS..134....1V} and \citet{1992ApJS...80..109B}.
The agreement is good.
We notice in particular that the predicted optical / UV ratio,
which in the simulations by \citet{2004ApJ...615..610B} is always too small
when both inward and outward \feii\ emission is considered, is
actually larger than observed (i.e., the UV \feii\ strength is underpredicted).
This allows for a component of UV \feii\ emission to form in
the broad-line region (BLR), as suggested by reverberation results of NGC 5548
\citep{1993ApJ...404..576M}.

This is a major advance in the comparison between
theory and observations for the \feii\ emission.
It is important to keep in mind the fact that the deduced
properties of emission-line clouds have, until now, come from
comparing the total emission, not the outward emission,
with observations.
Future papers will further
investigate the implications of asymmetric anisotropic emission.

\section{Discussion}

Equation (\ref{eqn:ColDenEddRatio}) suggests a
linear relationship
between the Eddington ratio and the column density of
infalling clouds.
Figure \ref{fig:VaryColumnDensity} shows that the
\feii/\hb\ ratio increases with increasing column density
for $N(\mathrm{H}) \sim 10^{22} - 10^{23} \pscm$.
This implies that there will be a relation between
$L / L_{Edd}$ and \feii/\hb .
This explains one of the strongest systematic correlations captured by
Eigenvector 1.  As was also recently suggested by \citet{dong09}, the cloud
column density is the physical variant which drives the spectroscopic
variations.

The predicted \feii/\hb\ reaches an asymptotic value of about 3
for large $N(\mathrm{H})$ or $L / L_{Edd}$.
This is close to the observed value for luminous quasars
\citep{1991ApJ...373..465F}.
Models which considered the full emission from the cloud
do not reproduce this ratio \citep{2004ApJ...615..610B}.

\la\ is strongly inwardly beamed,
while \civ\ is nearly isotropic,
as shown by \citet{1979ApJ...229..274F}
and \citet{1992ApJ...387...95F}.
\citet{2008ApJ...687...78H} find that all of the
emission in the optical \feii\ band originates in the infalling clouds.
Other lines are likely to form in both these and other regions,
including the accretion disk, virialized broad-lined clouds 
and outflowing winds.
High-resolution UV spectra of objects in which the infalling
clouds have the largest velocity shift are needed to test this.

The infalling clouds must have a substantial covering factor.
The observed \hb\ equivalent widths of the broad and infalling components
are similar \citep{2008ApJ...683L.115H}.
The infalling clouds direct much of their \hb\ emission
away from the observer.
They must have a covering factor approaching
$\Omega / 4 \pi \sim 0.5$, consistent with an origin
in the molecular torus
\citep{2008ApJ...687...78H}.

The net accretion rate due to these infalling clouds must be
\begin{equation}
\dot{M}_{\rm FeII} >
{\Omega\over 4\pi} 4\pi r^2 m_p (N_H/r) v_{\rm infall} \simeq
17 \,{\Omega\over 2\pi} r_{18} N_{\rm H,24} v_8
{\rm\,M_\odot\,yr^{-1}},
\end{equation}
where $r=10^{18}r_{18}{\rm\,cm}$, $N_{\rm H} = 10^{24}N_{\rm
H,24}{\rm\,cm^{-2}}$, and $v=1000v_8{\rm\,km\,s^{-1}}$.  In this
expression, the factor $N_H/r$ is the mean density of the
Fe{\sc\,ii}-emitting material multiplied by the minimal line-of-sight
filling factor if the clouds are evenly distributed.
This may be compared to the accretion rate of $1.8 L_{45}
(\eta/0.01)^{-1}{\rm\,M_\odot\,yr^{-1}}$ required to power an AGN of
luminosity $L = 10^{45}L_{45}{\rm\,erg\,s^{-1}}$ with an accretion
efficiency of $\eta$, where the scaling parameters are all $\simeq 1$
for Akn 120.  The mass infall rate above that required by accretion is
likely to be balanced by other outflows.  Nevertheless, this suggests
that the infalling \feii-emitting material constitutes a major element
in the mass budget of high Eddington-ratio active nuclei.

We have investigated the effects of 
systematically viewing a low-ionization cloud 
from its shielded face. 
We focus on 
\feii\ and \hb\ because of their importance in PCA analysis.
Meaningful predictions about 
the full spectrum will require a much broader exploration of, 
and probably a weighted integration over, the ionizing flux versus
gas density plane. 
A future paper will carry out that broader exploration and 
address predicted strengths and profiles of such lines, 
and also the open question of whether this infalling component 
should be seen in absorption particularly at UV or X-ray wavelengths. 

\section{Conclusions}
\label{sec:Conclusions}

\begin{itemize}

\item
Low-ionization \feii -emitting clouds
have force multipliers (the ratio of total to
electron scattering opacities) that are large if the clouds are optically thin.
Such clouds will be accelerated away from the central black hole
if it has an Eddington ratio as large as those found
in quasars, $L / L_{Edd} \sim 0.1$.

\item
If the \feii -emitting clouds are \emph{infalling}, 
as is suggested by their systemically positive velocity offsets
\citep{2008ApJ...687...78H}, they must have column densities 
significantly larger
than previously thought if the inward pull of gravity
is to offset the outward force of radiation pressure.

\item
There is a simple relationship between $L / L_{Edd}$
and the minimum cloud column density required for infall.
\feii/\hb\ also depends on cloud column density.
Cloud column density is the underlying
driver that couples Eigenvector 1, $L / L_{Edd}$,
and spectroscopic variations, as was previously
proposed by \citet{dong09}.

\item
Previous simulations of the emitting gas have assumed
a symmetric distribution of clouds so that, on average,
we see the same number of clouds from
their illuminated as from their shielded faces. 
We have investigated here the result of seeing an 
asymmetric distribution of emitters,
so that we mainly observe \feii\ emission from the
shielded face of near-side infalling clouds, 
as is suggested by observations (\citealp{2008ApJ...687...78H}, 
\citealp{2009arXiv0908.0386G}).

\item
\feii\ UV emission is emitted less isotropically
than optical \feii\ lines.
Previous work, which assumed a symmetric distribution of
emitters, could not explain the larger-than-predicted
ratio of optical to UV \feii\ emission or the
\feii/\hb\ ratio.
The predicted emission ratios
from the shielded face are in good agreement with observations.
This is a major advance in understanding the nature
of \feii\ emission in AGN.

\item
It is likely that there is a distribution of
cloud column densities.
If so, clouds with column densities smaller than
that given in Equation (\ref{eqn:ColDenEddRatio})
will be radiatively accelerated outward while
those with larger column densities will fall
in, with a column density cutoff depending on the AGN's luminosity.
This would be the mechanism that accounts for the observed
\citep{2002ApJ...565...78B,dong09} coupling
between $L/L_{\rm Edd}$ and the column density and \feii/\hb\ ratio.
Both outflow and infall are observed in AGN.
Could this also be a natural consequence of a range of $N(\mathrm{H})$ in the surviving clouds?

\end{itemize}

\acknowledgments
\label{sec:Acknowledgments}
We thank the referee for very helpful comments. G.J.F. thanks NSF and NASA for support
(AST0607028, AST0908877, and ATFP07-0124), J.A.B. acknowledges NASA HST grant
AR-10932, and C.H. and J.M.W. acknowledge support from NSFC-10733010 and 10821061,
CAS-KJCX2-YW-T03, and 973 project (2009CB824800).

\bibliographystyle{apj}
\bibliography{main}

\label{lastpage}
\end{document}